\documentclass[sigconf,authorversion,nonacm]{acmart}

\usepackage{booktabs}
\usepackage{enumerate}
\usepackage{multirow}
\usepackage{listings}
\usepackage{quoting}
\quotingsetup{vskip=0pt}
\usepackage{microtype}
\usepackage{xspace}
\usepackage{enumitem}
\usepackage{natbib}
\usepackage{xcolor,colortbl,hhline}
\usepackage{amsthm}
\usepackage{amsmath,relsize}
\usepackage{framed}
\usepackage{diagbox}
\usepackage{subcaption}
\usepackage{graphicx}
\usepackage{adjustbox}

\definecolor{lgray}{gray}{0.8}

\theoremstyle{definition}
\newtheorem{definition}{Definition}

\newcolumntype{C}[1]{>{\centering\arraybackslash}p{#1}}
\newcolumntype{L}[1]{>{\raggedright\arraybackslash}p{#1}}
\newcolumntype{G}[1]{>{\columncolor{lgray}\raggedright\arraybackslash}p{#1}}

\iftrue
\newcommand{\josh}[1]{{\color{red}(Josh: #1)}}
\newcommand{\suma}[1]{{\color{purple}(Sumaya: #1)}}
\newcommand{\alfred}[1]{\textcolor{red}{[Alfred: #1]}}
\newcommand{\yuqi}[1]{\textcolor{orange}{[Yuqi: #1]}}
\else 
\newcommand{\josh}[1]{}
\newcommand{\suma}[1]{}
\newcommand{\alfred}[1]{}
\newcommand{\yuqi}[1]{}
\fi

\usepackage{hyphenat}
\newcommand{\approach}{\textsc{sceno\-RITA}\xspace}
\newcommand{\approachPlus}{$\scriptstyle scenoRITA^{+}$\xspace}
\newcommand{\approachMinus}{$\scriptstyle scenoRITA^{-}$\xspace}
\newcommand{\approachRand}{$\scriptstyle scenoRITA^{\scriptscriptstyle R}$\xspace}

\AtBeginDocument{%
  \providecommand\BibTeX{{%
    \normalfont B\kern-0.5em{\scshape i\kern-0.25em b}\kern-0.8em\TeX}}}
    
\newcounter{finding}[section]
\newenvironment{finding}{\refstepcounter{finding}
\vspace{-1mm}
\framed
\noindent \textbf{Finding \thefinding:}
}
{
\endframed
\vspace{-1mm}
}

\usepackage{titlesec}
\titlespacing\section{0pt}{2pt plus 0pt minus 2pt}{0pt plus 0pt minus 1pt}
\titlespacing\subsection{0pt}{0pt plus 1pt minus 2pt}{0pt plus 0pt minus 2pt}
\titlespacing\subsubsection{0pt}{2pt plus 0pt minus 2pt}{4pt plus 2pt minus 1pt}

\expandafter\def\expandafter\normalsize\expandafter{%
    \normalsize
    \setlength\abovedisplayskip{3pt}
    \setlength\belowdisplayskip{2pt}
    \setlength\abovedisplayshortskip{5pt}
    \setlength\belowdisplayshortskip{5pt}
}

\makeatletter
\def\thm@space@setup{%
  \thm@preskip=0pt
  \thm@postskip=\thm@preskip 
}
\makeatother

\begin{document}

\title{\approach: Generating Less-Redundant, Safety-Critical and Motion Sickness-Inducing Scenarios for Autonomous Vehicles}

\author{Sumaya Almanee, Xiafa Wu, Yuqi Huai, Qi Alfred Chen and Joshua Garcia}
\affiliation{%
   \institution{University of California, Irvine}
   \city{}
   \state{}
   \country{}}
\email{{ salmanee, xiafaw, yhuai, alfchen, joshug4 } @uci.edu} 

\begin{abstract}
There is tremendous global enthusiasm for research, development, and deployment of autonomous vehicles (AVs), e.g., self-driving taxis and trucks from Waymo and Baidu. The current practice for testing AVs uses virtual tests---where AVs are tested in software simulations---since they offer a more efficient and safer alternative compared to field operational tests. Specifically, search-based approaches are used to find particularly critical situations. These approaches provide an opportunity to automatically generate tests; however, systematically creating \textit{valid} and \textit{effective} tests for AV software remains a major challenge. To address this challenge, we introduce \approach, a test generation approach for AVs that uses evolutionary algorithms with (1) a novel gene representation that allows obstacles to be \textit{fully mutable}, hence, resulting in more reported violations, (2) 5 test oracles to determine both safety and motion sickness-inducing violations, and (3) a novel technique to identify and eliminate duplicate tests. Our extensive evaluation shows that \approach can produce effective driving scenarios that expose an ego car to safety critical situations. \approach generated tests that resulted in a total of 1,026 \textit{unique} violations, increasing the number of reported violations by 23.47\% and 24.21\% compared to random test generation and state-of-the-art partially-mutable test generation, respectively.
\end{abstract}

\maketitle
\pagestyle{plain}

\section{Introduction} \label{sec:intro}
Autonomous vehicles (AVs), a.k.a. self-driving cars, are becoming a pervasive and ubiquitous part of our daily life. More than 50 corporations are
actively working on AVs, including large companies such as Google's parent company Alphabet,
Ford, and Intel~\cite{waymo,av_ford,av_intel}. Some of these companies (e.g., Alphabet's Waymo, Lyft, and Baidu) are already serving customers on public roads~\cite{waymo_public,lyft_public,baidu_apolong}. Experts forecast that AVs will drastically impact society, particularly by reducing accidents~\cite{expertAVSaftey}. However, crashes caused by AVs indicate that achieving this lofty goal remains an open challenge.
Despite the fact that companies such as Tesla~\cite{tesla}, Waymo~\cite{waymo}, or Uber~\cite{uber} have released prototypes of AVs with a high level of autonomy, they have caused injuries or even fatal accidents to pedestrians. For instance, an AV of Uber killed a
pedestrian in Arizona back in 2018 \cite{uber_crash}. AVs with lower levels of autonomy have resulted in more fatalities in recent years \cite{tesla_autopilot_crashes,tesla_crash_father_son,tesla_driver_death,nytimes_tesla_crash,tesla_mv_crash,waymo_crash,tesla_crash,uber_crash}.

Prior research has revealed a lack of standardized procedures to test AVs~\cite{standardTest} and the inability of current approaches to effectively translate traditional software testing approaches into the space of AVs~\cite{test_challenges_1,test_challenges_2}. A common practice for testing AV software lies in field operational tests, in which AVs are left to drive freely in the physical world. 
This approach is not only expensive and dangerous, but also ineffective since it misses critical testing scenarios~\cite{fot}.
Virtual tests, where AVs are tested in software simulations, offer a far more efficient and safer alternative. While these tests provide an opportunity to automatically generate tests, they come with the key challenge of \textit{systematically generating scenarios which expose AVs to safety-critical and motion sickness-inducing situations}.

To address this challenge, we propose \approach (\textbf{scen}ari\textbf{o} Gene\textbf{R}at\textbf{I}on \textbf{T}esting for \textbf{A}Vs), a test generation framework which aims to find safety and motion sickness-inducing violations in the presence of an evolving traffic environment. \approach combines both (i) AV software domain knowledge and (ii) search-based testing~\cite{search_based_survey,evosuite}. These two elements have been combined by previous techniques to test AVs by automatically generating safety-critical scenarios~\cite{avfuzzer,asfault,avoidCollision,althoffIV,ben2016testing,abdessalem2018testing}. However, unlike these approaches, \approach's gene representation enables obstacles to be \textit{fully mutable}, i.e., an obstacle's \textit{individual} properties such as its start and end location, type (e.g., vehicle, pedestrian, and bike), heading, speed, size, and mobility (e.g., static or dynamic) can be altered. While previous techniques do not specify their gene representations or do so in such a way that allows obstacles to be only \textit{partially mutable}, obstacles' attributes are altered only during mutation and with a small probability, while during crossover, obstacles are transferred across scenarios without altering their states or properties ~\cite{avfuzzer,ac3r,asfault,avoidCollision}. 
Thus, these techniques ignore the challenge of ensuring creation of valid obstacle trajectories, reducing their \textit{effectiveness at generating driving scenarios with unique violations}. 

Additionally, previous work on AV software testing uses a highly limited number of test oracles for ensuring safety and no oracles for assessing motion sickness-inducing movement of an AV: State-of-the-art AV testing approaches (\textit{AC3R}, \textit{AV-Fuzzer}, \textit{AsFault} and Abdessalem et al. \cite{ben2016testing,abdessalem2018testing}) use only two oracles for checking if (1) the ego car reaches its final expected position while avoiding a crash (i.e., collision detection) and (2) if a vehicle drives off the road (i.e., off-road detection). As a result, the fitness functions these techniques utilize are overly simplified---substantially reducing their degree of safety assurance while completely ignoring rider comfort and motion sickness.  Research has shown that a rider's discomfort increases when a human is a passenger rather than a driver---with up to one-third of Americans experiencing motion sickness, according to the National Institutes of Health (NIH)~\cite{nyt_motion_sickness,michigan_motion_sickness,leung2019motion}.

To overcome such limitations, \approach utilizes 5 test oracles (i.e., collision detection, speeding detection, unsafe lane change, fast acceleration, and hard braking) and corresponding fitness functions---which are based on grading metrics for driving behavior defined by Apollo's developers~\cite{grading_metrics}. Apollo is a high autonomy (i.e., Level 4), open-source, production-grade AV software system created by Baidu. The Society of Automotive Engineers (SAE) defines 6 levels of vehicle autonomy~\cite{sae2014taxonomy,level4}, where Level 4 (L4) AV systems, such as Apollo, have the AV perform all driving functionality under certain circumstances, although human override is still an option. Apollo is selected by Udacity to teach state-of-the-art AV technology~\cite{udacity-apollo} and can be directly deployed on real-world AVs such as Lincoln MKZ, Lexus RX 450h, GAC GE3, and others~\cite{baidu_apollo,apolloCars}, and has mass production agreements with Volvo and Ford~\cite{baidu_volvo_ford}. Additionally, Apollo has already started serving the general public in cities (e.g., a robo-taxi service in Changsha, China \cite{robotaxi}). 

\noindent The main contributions of this paper are as follows:
\begin{itemize}[leftmargin=*,nosep]
    \item We introduce \approach, a search-based testing framework, with a novel gene representation and \textit{domain-specific constraints}, that automatically generates \textit{valid} and \textit{effective} driving scenarios. \approach aims to maximize the number of scenarios with unique violations and relies on a novel gene representation of driving scenarios, which enables the search to be more \textit{effective}: Our gene representation allows the genetic algorithm to alter the states and properties of obstacles in a scenario, allowing them to be fully mutable. Additionally, we specify a set of \textit{domain-specific constraints} to ensure that the generated driving scenarios are \textit{valid}. To the best of our knowledge, we are the first to define the exact values of these constraints, which are obtained from authoritative sources such as the National Center for Health Statistics, Federal Highway Administration, and the US Department of Transportation~\cite{human_measurement,bike_measurement_speed,commercial_cars_measurement_speed,lanechange_duration}.
    \item To improve the effectiveness of \approach, we automate the process of identifying and eliminating duplicate violations by using an unsupervised clustering technique to group driving scenarios, with similar violations, according to specific features.
    \item We utilize 5 test oracles and corresponding fitness functions to assess different aspects of AVs---ranging from traffic and road safety (i.e., collision detection, speeding detection, unsafe lane change) to a rider's comfort (i.e., fast acceleration and hard braking). To the best of our knowledge, \approach is the first search-based testing technique for AV software that uses multiple test oracles at the same time and considers both comfort and safety violations as part of those oracles.
    \item We evaluate the efficiency and effectiveness of \approach by comparing it with random search and a state-of-the art search-based approach---which adopts the gene representation in prior work~\cite{avfuzzer,avoidCollision,asfault} that allows \textit{only} partial mutation of obstacles (i.e., obstacles are transferred across scenarios without altering their states or properties during crossover). 
\end{itemize}

\noindent Our extensive evaluation---which consists of executing a total 
of 31,413 virtual tests on Baidu Apollo using 3 high-definition maps of cities/street blocks located in California: Sunnyvale (3,061 lanes); San Mateo (1,305 lanes); and Borregas Ave (60 lanes)---shows that \approach generates driving scenarios that expose the ego car to critical and realistic situations.  \approach found 1,026 \textit{unique} comfort and safety violations, while random testing and the partially mutable search-based testing found a total of 826 and 831, respectively. We make our testing platform, dataset and results available online to enable reusability, reproducibility, and others to build upon our work~\cite{scenorita-repo}.

\section{Specification of the State Space}\label{formal_spec}
To aid in the generation of effective and valid scenarios, we present a formal specification of the state space in the form of driving scenarios. \approach uses this formal specification of the state space, along with a genetic algorithm, to generate scenarios that maximizes the possibility of the ego car (i.e., AV) either violating safety or causing rider discomfort. 

\theoremstyle{definition}
\begin{definition}\label{def:scenario}
A Scenario $\mathcal{S}_c$ is a tuple $\langle t, E, \mathbb{O}, \mathbb{L} \rangle$ where:
\begin{itemize}[leftmargin=*,nosep]
    \item t is a finite number that represents the maximum duration of $\mathcal{S}_c$. 
    \item $E$ is the only ego car (i.e., the autonomous driving car) in $\mathcal{S}_c$.
    \item $\mathbb{O}$ is a finite, non-empty set of $n$ obstacles (i.e. non-player characters). A \textit{single} obstacle is represented as $O_k$ where $\mathbb{O}= \{O_k: 1 \leq k \leq n \}$.
    \item $\mathbb{L}$ is a finite, non-empty set of lanes, where $E$ and $\mathbb{O}$ reside/travel. 
\end{itemize}
\end{definition}

\theoremstyle{definition}
\begin{definition}\label{def:ego}
An ego car $E$, is a tuple $\langle Z_E, \mathbb{H}_{E},\mathbb{P}_E,\mathbb{S}_E,\mathbb{A}_E,\mathbb{C}_E \rangle$ where:
\begin{itemize}[leftmargin=*,nosep]
     \item $Z_{E} = \langle wid,len,hgt \rangle$ represents the width $wid$, length $len$, and height $hgt$ of the ego car $E$.
      \item $\mathbb{H}_{E}$ is a finite, non-empty set representing the ego car's headings during time instants of $\mathcal{S}_c$. The heading of $E$ at timestamp $j$ is represented as $h_{j}^{E}$ where $\mathbb{H}_{E}= \{h_{j}^{E}: 1 \leq j \leq t \}$. 
     \item $\mathbb{P}_{E}$ is a finite, non-empty set representing the ego car's positions during time instants of $\mathcal{S}_c$. The position of $E$ at timestamp $j$ is represented as $p_{j}^{E}$ where $\mathbb{P}_{E}= \{p_{j}^{E}: 1 \leq j \leq t \}$. 
     \item $\mathbb{S}_{E}$ is a finite, non-empty set representing the ego car's speed during time instants of $\mathcal{S}_c$. The speed of $E$ at timestamp $j$ is represented as $s_{j}^{E}$ where $\mathbb{S}_{E}= \{s_{j}^{E}: 1 \leq j \leq t \}$.
    \item $\mathbb{A}_{E}$ is a finite, non-empty set representing the ego car's acceleration at time instants of $\mathcal{S}_c$. The acceleration of $E$ at timestamp $j$ is represented as $a_{j}^{E}$ where $\mathbb{A}_{E}= \{a_{j}^{E}: 1 \leq j \leq t \}$. 
    \item $\mathbb{C}_{E}$ is a finite, non-empty set of durations an ego car spends driving at the boundary of two lanes at the same time. When an ego car changes lanes, it drives on the markings between two lanes for a period of time $c$, before it completely switches to the target lane. The duration $E$ spends driving on the markings at timestamp $j$ is represented as $c_{j}^{E}$ where $\mathbb{C}_{E}= \{c_{j}^{E}: 1 \leq j \leq t \}$.
\end{itemize}
\end{definition}

\theoremstyle{definition}
\begin{definition}\label{def:obstacle} 
A \textit{single} obstacle $O_k$ in $\mathcal{S}_c$ is a tuple \\
$\langle ID_{O_k}, T_{O_k}, Z_{O_k}, M_{O_k}, \mathbb{H}_{O_k}, \mathbb{P}_{O_k}, \mathbb{S}_{O_k} \rangle$ where:
\begin{itemize}[leftmargin=*,nosep]
    \item $ID_{O_k}$ represents a unique identification number associated with $O_k$.
     \item $T_{O_k}$ represents the type of an obstacle. Examples of obstacle types are: \texttt{VEHICLE, PEDESTRIAN, and BICYCLE}.
     \item $Z_{O_k} = \langle wid,len,hgt \rangle$ represents the width $wid$, length $len$, and height $hgt$ of obstacle $O_k$.
     \item $M_{O_k}$ represents the mobility of an obstacle (e.g., static or mobile).
     \item $\mathbb{H}_{O_k}$ is a finite, non-empty set representing the direction of $O_k$ during the entire duration of $\mathcal{S}_c$. The heading of $O_k$ at timestamp $j$ is represented as $h_{j}^{O_k}$ where $\mathbb{H}_{O_k}= \{h_{j}^{O_k}: 1 \leq j \leq t \}$.
     \item $\mathbb{P}_{O_k}$ is a finite, non-empty set representing the positions of $O_k$ at time instants of $\mathcal{S}_c$. The position of $O_k$ at timestamp $j$ is represented as $p_{j}^{O_k}$ where $\mathbb{P}_{O_k}= \{p_{j}^{O_k}: 1 \leq j \leq t \}$.
       \item $\mathbb{S}_{O_k}$ is a finite, non-empty set representing the speed of $O_k$ at time instants of $\mathcal{S}_c$. The speed of $O_k$ at timestamp $j$ is represented as $s_{j}^{O_k}$ where $\mathbb{S}_{O_k}= \{s_{j}^{O_k}: 1 \leq j \leq t \}$.
\end{itemize}
\end{definition}

\theoremstyle{definition}
\begin{definition}\label{def:lanes}
A single lane $l \in \mathbb{L}$ has a speed limit $\mathbb{S}_l$ , which is a finite, non-empty set representing the speed limit imposed by $l$. The speed limit of $l$, which the ego car is traversing at timestamp $j$, is represented as $s_{j}^{l}$ where $\mathbb{S}_{l}= \{s_{j}^{l}: 1 \leq j \leq t \}$.
\end{definition}

\theoremstyle{definition}
\begin{definition}\label{def:violations}
We define a violation $v \in \mathbb{V}=\{collision, speed, \\ \mathit{unsafeChange}, \mathit{fastAccl}, hardBrake\}$. We elaborate on the oracles corresponding to each of these violations in Section \ref{sec:scenorita:gmc}.
\end{definition}

\section{\approach}\label{approach}
\begin{figure}[ht]
\includegraphics[width=0.96\linewidth]{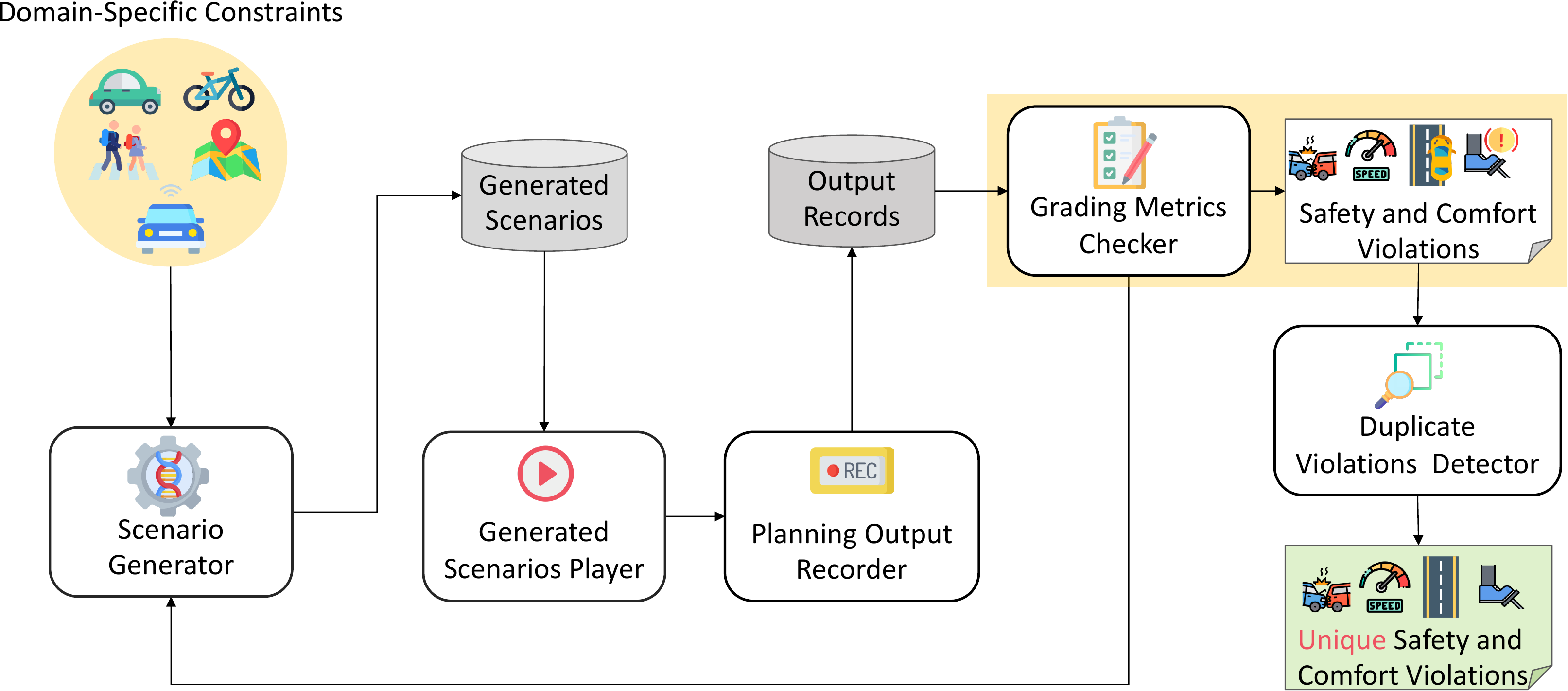}
\caption{An Overview of \approach} 
\label{fig:overview}
\end{figure}

\autoref{fig:overview} shows the overall workflow of \approach. Our main goal is to create valid and effective driving scenarios that expose AV software to unique safety and comfort violations.  \approach achieves this goal as follows: (1) it takes as an input a set of \textit{domain-specific constraints}, which dictates what constitutes a \textit{valid} driving scenario (e.g., obstacles should be moving in the direction of traffic in the lane and have valid obstacle identifiers); (2) The \textit{Scenario Generator} uses a genetic algorithm to produce driving scenarios with randomly generated but valid  obstacles, following the \textit{domain-specific constraints}. The genetic algorithm evolves the driving scenarios with the aim of finding scenarios with safety and comfort violations; (3) \textit{Generated Scenarios Player} converts the genetic representation of scenarios (\textit{Generated Scenarios}), from the previous step, into driving simulations where the planning output of the AV under test is produced and recorded by \textit{Planning Output Recorder};
(4) The planning output is then evaluated by \textit{Grading Metrics Checker} for \textit{safety and comfort violations}; (5) When the evolution process terminates, the \textit{Duplicate Violations Detector} inspects the violations produced by \textit{Grading Metrics Checker} to eliminate any duplicate violations, and produces a set of \textit{unique safety and comfort violations}. In the remainder of this section, we discuss each of these elements of \approach in more detail.

\subsection{Domain-Specific Constraints}\label{ds_constraints}
\textit{Domain-specific constraints} specifies the constraints that \textit{Scenario Generator} should follow to ensure that the generated driving scenarios are \textit{valid}. In this work, we define \textit{valid scenarios} as those which contain obstacles that are (1) moving in the direction of traffic in the lane; (2) having start and end points contained within the boundaries of a fixed-size map; and (3) having dimensions (i.e. width, height, and length) and speed that account for the obstacle type (vehicle, pedestrian, or bike). For example, the speed of a pedestrian should not exceed the average walking/running speed of a human. 

When generating the initial and final position of the ego car and obstacles, \textit{Scenario Generator} must ensure that these points are (i) within the boundaries of a fixed-size map, and (ii) have  a valid path allowing the ego car and obstacles to move in the direction of traffic. 
To ensure that an obstacle is driving in the right direction, the heading attribute should be within a valid range. An obstacle heading is the compass direction in which an obstacle is pointing at a given time, and it is expressed as the angular distance relative to the north. An obstacle's heading has a positive value in the counter-clockwise direction, with a maximum of  $180^\circ$ ($\pi$ or $3.14rad$), and a negative value in the clockwise direction, with a minimum of $-180^\circ$ ($-\pi$ or $-3.14rad$). 
 
Unlike prior work, we consider a wide range of obstacle-related attributes including type, size, speed and mobility. An obstacle can be a \texttt{VEHICLE}, \texttt{BICYCLE}, or a \texttt{PEDESTRIAN}, and the type of an obstacle dictates the minimum and maximum allowed values of its size and speed. An obstacle is represented as a polygon, hence, its size is expressed in terms of the width, height, and length (i.e., $Z_{O_k}$) of the polygon. The ego car $E$ is similarly represented as a polygon based on $Z_E$.

To determine the maximum and minimum dimensions of a pedestrian, we followed the most recent report published by the National Center for Health Statistics (NCHS)~\cite{human_measurement}, which provides the most recent anthropometric reference data for children and adults in the United States.
The height of a pedestrian ranges from 0.97m (average height of a child) to 1.87m (average height of an adult aged 20+). The width (shoulder width) ranges from 0.24m to 0.67m, while the length ranges from 0.2m to 0.45m. 
The speed of a pedestrian can range from 4.5km/hr (average walking speed) to 10.5 km/hr (average running speed)~\cite{human_speed}.

To determine the maximum and minimum dimensions and speed for both bicycles and vehicles, we followed the size and speed regulations imposed by the Federal Highway Administration and the US Department of Transportation~\cite{bike_measurement_speed,commercial_cars_measurement_speed}. The speed of a bicycle can range from 6km/hr to 30 km/hr, while the speed of a vehicle can range from 8km/hr (e.g., parking lots) to 110km/hr (e.g., highways). 

\subsection{Scenario Generator}\label{scenario_gen}
Our overarching goal is to create valid and effective driving scenarios that expose AV software to safety and comfort violations. The \textit{Scenario Generator} takes as input a set of \textit{domain-specific constraints} (\autoref{ds_constraints}) and uses a genetic algorithm to maximize a defined set of fitness functions (representing safety and comfort violations) to guide the search for problematic scenarios. The genetic algorithm is initialized with a starting population of tests (i.e., driving scenarios). To evaluate the fitness of tests, scenario representations are transformed into driving simulations, in which navigation plans are generated based on the origin and destination of the ego car. Additionally, driving trajectories/plans are computed for the ego car based on the scenario set-up (e.g., number of obstacles, state of the obstacles, ego car start and target position, etc.). During the simulation, the driving decisions of the ego car (e.g., driving maneuvers, stop/yield decisions, acceleration) are recorded by \textit{Planning Output Recorder} at regular intervals to identify safety and comfort violations. 
A set of values such as the distance between the ego car and other obstacles, the distance between the ego car and lane boundaries, the speed of the ego car, the acceleration and deceleration of the ego car are used to compute the fitness of individuals (\autoref{fitness}). These values guide the genetic algorithm when evolving test cases (i.e., driving scenarios) by recombining and mutating their attributes (\autoref{search_op}). The algorithm continues to execute and evolve test cases until a user-defined ending condition is met, at that point \approach returns the final test suite and stops.


\subsubsection{Representation.}
A set of individuals together represent a driving scenario which, in turn, represents a single test. A test suite in \approach is a set of driving scenarios. An individual is represented as a vector, where each index in the vector represents a gene. The number of input genes is fixed, where the 10 genes corresponds to the following 10 attributes of a \textit{single} obstacle $O_k$: $ID_{O_k}$; the initial position $p_1^{O_k}$ of $O_k$; the final position $p_t^{O_k}$ of $O_k$ at the final timestamp $t$; heading $h_j^{O_k}$ at timestamp $j$; length $Z_{O_k}.len$; width $Z_{O_k}.wid$; and height $Z_{O_k}.hgt$; speed $s_j^{O_k}$ at timestamp $j$; type $T_{O_k}$; and mobility $M_{O_k}$. Each gene value can change (e.g., when initialized or mutated), but it still has to adhere to the valid ranges defined in \autoref{ds_constraints}. 

Representing \textit{obstacles} as \textit{individuals} allows \approach to alter obstacles' attributes and states when applying search operators, hence, allowing obstacles to be \textit{fully mutable} throughout the test generation process. \autoref{fig:cx}(b)  demonstrates \approach's application of a crossover operator on two individuals (i.e., obstacles) compared to how related work recombine their individuals (\autoref{fig:cx}(c)). Previous approaches \cite{asfault,avoidCollision,avfuzzer}, represent \textit{obstacles} as \textit{genes}, resulting in obstacles being \textit{partially mutable} during recombination and mutation operators. For example, the crossover operator in \textit{AV-Fuzzer}~\cite{avfuzzer} does not alter properties of obstacles, instead it simply swaps two
randomly selected obstacles in two scenarios, \textit{Scenario B} and \textit{Scenario D}, with a certain probability. For the remainder of the paper, we will use \approachPlus and \approachMinus to refer to \textit{fully mutable} and \textit{partially mutable} approaches, respectively. 

By representing obstacles using individual attributes (e.g., location), as opposed to just as a gene in the case of \approachMinus, \approachPlus must address the challenge of ensuring the creation of valid obstacle trajectories. To that end, \approachPlus includes logic that allows it to check whether there is a valid path between the newly-generated start and end locations of an obstacle. If the recombination operator introduces an invalid path, then \approachPlus generate new locations for an obstacle until a valid one is found. This process allows \approachPlus to represent an obstacle's individual attributes, such as the start and end locations, as genes while preventing the invalid genes that may be produced due to the application of search operators. Our evaluation results in \autoref{sec:eval} shows that \approachPlus, which corresponds to our proposed gene representation, results in 23.47\% more effective and unique violations compared to \approachMinus.

\subsubsection{Fitness Evaluation.}\label{fitness} 
In each generation, individuals are assessed for their fitness, with respect to the search objective, to be selected to pass on their genes. \approach determines the fitness of an individual by evaluating how close they are in terms of causing safety or comfort violations. This is measured by calculating an individual $i$'s fitness using a function $f_v(i)$ with respect to a safety and comfort violation $v$. Recall from \autoref{sec:intro} that \approach considers 5 safety and comfort violations based on the grading metrics defined by Apollo's developers \cite{grading_metrics} and, thus, represents violation constructs and thresholds used by professional AV developers. Three of these metrics assess driving scenarios for traffic and road safety (collision detection, speeding detection, and unsafe lane change), while the remaining two metrics assess a rider's comfort (fast acceleration and hard braking). 

The fitness of an individual $i$ is determined as follows:
\begin{equation}\label{eq:general}
F(i)=\sum_{v \in \mathbb{V}} f_{v}(i)
\end{equation}

\noindent where $\sum_{v \in \mathbb{V}} f_{v}(i)$ determines the fitness of an individual with respect to all 5 safety and comfort violations. Recall that $v \in \mathbb{V}=\allowbreak \{collision,\allowbreak speed,\allowbreak \mathit{unsafeChange},\allowbreak \mathit{fastAccl},\allowbreak hardBrake\}$ (\autoref{def:violations}), hence, $F(i)$ aims to maximize the number of safety violations. In the remainder of this section, we define $f_{v}(i)$ in more detail. 

\noindent\textbf{Collision Detection}. In the context of collision detection, effective tests are those which cause the ego car to collide with other obstacles. Therefore, \approach uses as a fitness function $f_{collision}$ (\autoref{eq:collision}), which rewards tests that cause the ego car
to move as close as possible to other obstacles. Given a simulated scenario with
a maximum duration of $t$, a set of positions $\mathbb{P}_E$ for the ego car $E$, and a set of positions for the $k^{th}$ obstacle in a scenario defined as  $\mathbb{P}_{O_k}= (p_{1}^{O_k} , p_{2}^{O_k} , ... , p_{t}^{O_k}) $, we define $f_{collision}$ as:
\begin{equation}
\label{eq:collision}
\begin{split}
f_{collision}(i)=&\, \min\{ D_c(p_{j}^{E},p_{j}^{O_k}): 1 \leq j \leq t, \\ 
&\, p_{j}^{E} \in \mathbb{P}_E, p_{j}^{O_k} \in \mathbb{P}_{O_k}\}
\end{split}
\end{equation}

\noindent $D_c(p^{E},p^{O_k})$ is the shortest distance between the position of a given obstacle and the ego car at a given time. 
$f_{collision}$ captures the intuition that tests causing the ego car
to drive closer to other obstacles (i.e., have a minimum distance between the ego car and an obstacle) are more likely to lead to a collision.
\vspace{2mm}

\noindent\textbf{Speeding Detection}. We use a fitness function $f_{speed}$ (\autoref{eq:speed}), which rewards tests that cause the ego car to exceed the speed limit of the current lane. Given a simulated scenario with a maximum duration $t$, the speed profile $\mathbb{S}_{E}$ of the ego car $E$, and a set of speed-limits imposed by lanes of which the ego car traversed $\mathbb{S}_{L}$, we define $f_{speed}$ as:
\begin{equation}\label{eq:speed}
f_{speed}(i)= \min\{ D_s(s_{j}^{l},s_{j}^{E}) : 1 \leq j \leq t, s_{j}^{E} \in \mathbb{S}_{E},  s_{j}^{l} \in \mathbb{S}_{L}\}
\end{equation}

\noindent $s_{j}^{E}$ and $s_{j}^{l}$ represent the ego car speed and the speed limit of the lane in which the ego car is travelling at timestamp $j$. Furthermore, $D_s(s^{l},s^{E})$ is the difference between the speed limit imposed by a given lane  and the current speed of the ego car. 
$f_{speed}$ captures the intuition that as the ego car approaches the speed limit of a given lane it is more likely to result in speed violations. 
\vspace{2mm}

\noindent\textbf{Unsafe Lane Change}. A lane change is defined as a driving maneuver that moves a vehicle from one lane to another, where both lanes have the same direction of travel. We primarily focus on the \textit{duration} the ego car spends travelling at the boundary of two lanes while changing lanes. We define a \textit{safe} lane-change duration as $\delta_{safe}$. 
We define a fitness function $f_{\mathit{unsafeChange}}$ (\autoref{eq:unsafelc}), which rewards tests that cause the ego car to spend more than $\delta_{safe}$ driving at the boundary of two lanes. Given a simulated scenario with lane-change durations $\mathbb{C}_{E}$ for ego car $E$, and a safe lane-change duration $\delta_{safe}$, we define $f_{\mathit{unsafeChange}}$ as:\vspace{-1mm}
\begin{equation}\label{eq:unsafelc}
f_{\mathit{unsafeChange}}(i)=\max(c_{j}^{E})
\end{equation}
\noindent $c_{j}^{E} \in\mathbb{C}_{E}$  represents the duration an ego car spends driving between two lanes. $f_{\mathit{unsafeChange}}$ captures the intuition that tests that cause the ego car to spend longer periods of time driving on lane boundaries are more likely to result in an unsafe lane change violation.

\noindent\textbf{Fast Acceleration}. We use a fitness function $f_{fastAccl}$ (\autoref{eq:fastaccl}), which rewards tests that cause the ego car to accelerate too fast, potentially inducing motion sickness. Given a simulated scenario with a maximum duration $t$ and the acceleration profile $\mathbb{A}_{E}$ for the ego car $E$, we define $f_{fastAccl}$ as:
\begin{equation}\label{eq:fastaccl}
f_{fastAccl}(i)=\max\{a_{j}^{E}: 1 \leq j \leq t, a_{j}^{E} \in \mathbb{A}_{E} \}
\end{equation}
$a_{j}^{E} \in \mathbb{A}_{E}$ represents the acceleration of the ego car $E$ at timestamp $j$.  $f_{fastaccl}$ aims to maximize the acceleration of $E$ to induce a motion sickness violation. 

\noindent\textbf{Hard Braking}. We use a fitness function $f_{hardBrake}$ (\autoref{eq:hardbrake}), which rewards tests that cause the ego car to brake too hard (i.e., brake suddenly in a manner that induces motion sickness). Given a simulated scenario with a maximum duration $t$, the acceleration profile for the ego car defined as $\mathbb{A}_{E}$, we define $f_{hardBrake}$ as:
\begin{equation}\label{eq:hardbrake}
f_{hardBrake}(i)=\min\{a_{j}^{E}: 1 \leq j \leq t, a_{j}^{E} \in \mathbb{A}_{E} \}
\end{equation}
$a_{j}^{E} \in \mathbb{A}_{E}$ represents the deceleration of the ego car.  $f_{hardBrake}$ captures the intuition that tests which cause the ego car to decelerate too fast can result in hard braking.

\subsubsection{Search Operators.}\label{search_op}
\approach evolves driving scenarios by applying search operators, which mutate and recombine the scenario attributes according to certain probabilities. In this section, we provide a detailed explanation of these search operators.\vspace{1mm}

\noindent \textbf{Selection}. 
\approach uses the \textbf{N}on-dominated \textbf{S}orting \textbf{G}enetic \textbf{A}lgorithm selection (NSGA-II \cite{nsga2}) for breeding the next generation. NSGA-II is an effective algorithm used for solving multi-objective optimization problems (i.e., problems with multiple conflicting fitness functions) and further aims to maintain diversity of individuals. 

\begin{figure}[ht]
\includegraphics[width=0.99\linewidth]{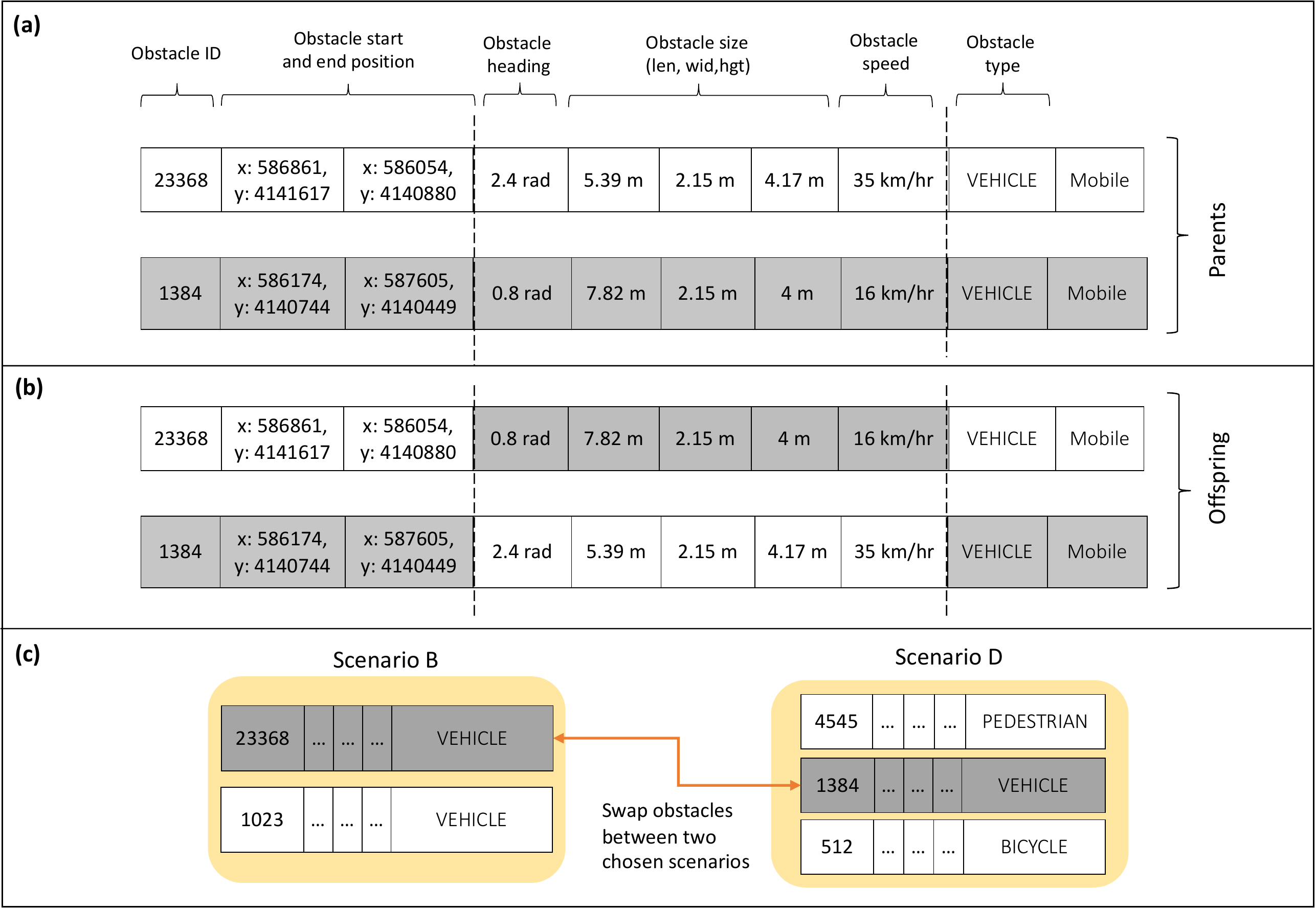}
\caption{(a) two individuals \textit{before} a crossover, (b) the same individuals \textit{after} a crossover for \approach, and (c) how crossover is applied in prior work \cite{avfuzzer}}.
\label{fig:cx}
\end{figure}

\noindent \textbf{Crossover}. 
This operator selects two individuals from a given scenario and creates superior offspring by mixing their parents' genetic makeup. \approach uses a two-point crossover strategy, where two crossover points are picked randomly from the mating individuals (i.e., parents) and the genes between the two points are swapped. \autoref{fig:cx} illustrates the application of the two-point crossover operator on two sample individuals; the two individuals are modified in place and both keep their original length. We opt for the two-point crossover strategy since it maintains the length of individuals, in addition to increasing the extent of disruption in their original values.


Crossover may produce invalid scenario configurations. 
For example, after crossover of a vehicle's attributes with those of a pedestrian, some obstacle attributes produced in the offspring may violate the speed and size constraints in \autoref{ds_constraints}, such as having a pedestrian's speed changed from \textbf{10} km/hr (a valid running speed for a pedestrian) to \textbf{60} km/hr (an unrealistic speed for a pedestrian). When such a case is detected, \approach replaces the violated obstacle attributes with randomly generated values that fall within the valid ranges described in \autoref{ds_constraints}. 
\vspace{1mm}

\noindent \textbf{Mutation}. 
\approach applies the mutation operator to driving-scenarios in two forms: (1) it mutates individuals in a single scenario; and (2) it applies mutation operators across scenarios.  The first type of mutation, randomly replaces genes in individuals with new ones, where the newly generated values follow the constraints defined in \autoref{ds_constraints}. For example, the mutation operator can change the speed of a vehicle from \textbf{35} km/hr to \textbf{50} km/hr. This type of mutation does not change the number of individuals in a single scenario. The second type of mutation aims to diversify the \textit{number} of individuals in a given scenario $S$ by (1) adding the fittest obstacle from another randomly selected scenario $S_r$ to $S$, or (2) removing the least fit individual in $S$. 


\subsection{Generated Scenarios Player}\label{scenario_player}
Converting the genotypic representation of tests to driving simulations is the key task of \textit{Generated Scenarios Player}. Our approach uses code templates to generate the necessary simulation code, which instantiates the obstacles in the driving simulation, places the ego car in the starting position, and sets the target position of the ego car. Once the simulation code is ready, \textit{Generated Scenarios Player} starts the driving simulator, where the planning module computes the driving trajectory taking into consideration various aspects of the vehicle and elements of its environment (e.g., distance to lane center, smoothness of the trajectory).
\looseness-1

\subsection{Planning Output Recorder}\label{output_recorder}
The behavior of the ego car in the driving simulation (\autoref{scenario_player}) is recorded by the \textit{Planning Output Recorder}, which stores the ego car's driving behaviours in a record file. The recorded output enables \textit{Grading Metrics Checker} to identify and report the occurrences of \textit{safety and comfort violations}. These violations are checked when the driving scenario ends; hence, it does not halt the driving simulation after observing the first violation; instead, tests continue until the end of the scenario. This approach balances the cost of running expensive simulations with the benefit of collecting as many violations as possible. \approach uses the output records for reporting violations, and evaluating the fitness of tests which guides the evolution process in \autoref{scenario_gen}. Additionally, stored records enable us to replay scenarios with reported violations after \approach ends; this allows us to verify the correctness of generated tests (\autoref{eval:accruacy}), and to closely analyze them along with their underlying causes. We make these record files available for researchers and practitioners to reuse, replicate, or analyze them in future work~\cite{scenorita-repo}.

\subsection{Grading Metrics Checker} \label{sec:scenorita:gmc}
Previous work~\cite{avfuzzer,asfault,ac3r,ben2016testing,abdessalem2018testing,avoidCollision} considers a limited number of test oracles, mainly consisting of one test oracle per work (either collision detection or lane keeping). The limited use of test oracles found in such techniques ignores important safety and comfort issues (e.g., driving between lanes for too long or causing motion sickness) and provides significantly less insight into the testing of industry-grade AVs.
Unlike previous work, we consider 5 test oracles based on grading metrics defined by Apollo's developers \cite{grading_metrics}. These grading metrics test different aspects of AVs, ranging from traffic and road safety to a rider's comfort. In the remainder of this section, we describe each grading metric in detail along with the definition of its corresponding test oracle.

The \textbf{Collision Detection} oracle checks if the ego car reaches its final destination without colliding with other obstacles. The test oracle's passing condition (i.e., not resulting in a violation) for collision detection is defined as follows: 
\begin{equation}
\displaystyle\mathop{\mathlarger{\mathlarger{\mathlarger{\forall}}}}_{1 \leq j \leq t} D_c(p_{j}^{E},p_{j}^{O_k}) \neq 0
\end{equation}
\noindent where $t$ is the total duration of the scenario, and $D_c(p_{j}^{E},p_{j}^{O_k})$ is a function that calculates the \textit{shortest distance} between the position of the ego car $p_{j}^{E}$ and the position of the $k^{th}$ obstacle $p_{j}^{O_k}$ at timestamp $j$. The distance is measured, in meters, between the closest two points between the ego car's polygon and an obstacle polygon. If function $D_c$ returns a distance of \textbf{zero} meters between the ego car and any other obstacle, this indicates the occurrence of a collision.

The \textbf{Speeding Detection} oracle checks if the ego car reaches its final destination without exceeding the speed limit. The test oracle's passing condition (i.e., not resulting in a violation) for speeding detection is defined as follows: 
\begin{equation}
\displaystyle\mathop{\mathlarger{\mathlarger{\mathlarger{\forall}}}}_{1 \leq j \leq t} D_s(s_{j}^{l},s_{j}^{E}) \leq \beta_{safe}
\end{equation}
\noindent where $t$ is the total duration of the scenario, and $D_s(s_{j}^{l},s_{j}^{E})$ is a function that calculates the difference between the ego car's speed $s_{j}^{E}$ and the speed limit of the current lane $s_{j}^{l}$ at timestamp $j$. $\beta_{safe}$ represents the allowed threshold for an ego car to drive \textit{above} the current speed limit. We allow the ego car to exceed the current speed limit by a maximum of \textbf{8} km/hr, anything above that is considered a speed violation. We allow some degree of driving above the speed limit, since it can be unsafe for the ego car to drive below or at the speed limit in certain conditions~\cite{speed_limit_google} (e.g., driving at the speed limit when other cars are going much faster can be dangerous). 

The \textbf{Unsafe Lane-Change} oracle checks if the ego car reaches its final destination without exceeding a time limit $\delta_{safe}$ when changing lanes. Recall from \autoref{fitness}, that $\delta_{safe}$ represents a safe lane-change duration, which averages at 5 seconds \cite{lanechange_duration}. The test oracle's passing condition (i.e., not resulting in a violation) for unsafe lane change is defined as follows: 
\begin{equation}
\displaystyle\mathop{\mathlarger{\mathlarger{\mathlarger{\forall}}}}_{1 \leq j \leq t} c_{j}^{E} \leq \delta_{safe}
\end{equation}
\noindent where $t$ is the total duration of the scenario, and $c_{j}^{E}$ represents the duration an ego car spends driving between two lanes at timestamp $j$. If $c_{j}^{E}$ at a given time exceeds $\delta_{safe}$, this indicates the occurrence of an unsafe lane change.

The \textbf{Fast Acceleration} oracle checks if the ego car reaches its final destination without causing a rider's discomfort by accelerating too fast. The test oracle's passing condition for fast acceleration is defined as follows: 
\begin{equation}
\displaystyle\mathop{\mathlarger{\mathlarger{\mathlarger{\forall}}}}_{1 \leq j \leq t} a_{j}^{E} \leq \gamma_{\mathit{comfort}}
\end{equation}
\noindent where $t$ is the total duration of the scenario, and $a_{j}^{E}$ is the acceleration of the ego car at timestamp $j$. $\gamma_{comfort}$ represents the maximum acceleration allowed for an ego-car before it violates a rider's comfort. We allow the ego car to accelerate to a maximum of \textbf{4} $m/s^2$, a threshold utilized in prior work~\cite{comfort} and set by Apollo developers~\cite{grading_metrics}.

The \textbf{Hard Braking} oracle checks if the ego car reaches its final destination without causing a rider's discomfort by braking suddenly and excessively. The test oracle's passing condition for hard braking is defined as follows: 
\begin{equation}
\displaystyle\mathop{\mathlarger{\mathlarger{\mathlarger{\forall}}}}_{1 \leq j \leq t} a_{j}^{E} \geq \epsilon_{\mathit{comfort}}
\end{equation}
\noindent where $t$ is the total duration of the scenario, and $a_{j}^{E}$ is the acceleration of the ego car at timestamp $j$. $\epsilon_{comfort}$ represents the minimum acceleration allowed for an ego car before it violates a rider's comfort. We allow the ego car to decelerate to a minimum of \textbf{-4} $m/s^2$, a threshold used in prior work~\cite{comfort} and set by Apollo's developers~\cite{grading_metrics}.

\subsection{Duplicate Violations Detector}
\label{sec:scenorita:dvd}

One of the challenges of scenario-based testing is the possibility of producing driving scenarios with \textit{similar} violations. To improve the effectiveness of test generation, the \textit{Duplicate Violations Detector} automates the process of identifying and eliminating duplicate violations; it achieves this by using an unsupervised \textit{clustering} technique~\cite{dbscan} to group driving scenarios, with similar violations, according to specific \textit{features}. 

The set of features used by the clustering algorithm are extracted from the recorded files (\autoref{output_recorder}). For a \textbf{collision} violation, the \textit{Duplicate Violations Detector} extracts 8 features at time $t_c$, where $t_c$ indicates the first timestamp at which a collision occurs. These features include the location of the ego car $p^{E}_{t_c}$; ego car's speed $s^{E}_{t_c}$; ego car's heading $h^{E}_{t_c}$; $collision^{type}$, which indicates where a collision occurs in respect to the ego car (e.g., ``rear-end'', ``front'', ``left'', etc.); the type of the obstacle ($O_k$) that collided with the ego car $T_{O_k}$; the obstacle's size $Z_{O_k}$; obstacle's speed at collision time $s^{O_k}_{t_c}$; and obstacle's heading $h^{O_k}_{t_c}$. 

For the remaining violations, we extract their respective features at times $t_s$, $t_u$, $t_f$, and $t_h$, which correspond to the first timestamp a \textbf{speeding}, \textbf{unsafe lane change}, \textbf{fast acceleration}, and \textbf{hard braking} occurs, respectively. These features include the ego car $E$'s location at a violation time $p^{E}$, the speed $s^{E}$ of $E$, the heading $h^{E}$ of $E$, the length of time for which a violation lasts ($duration$), and the violation $value$.

the length of time for which a violation lasts before the ego car drives within the allowed safety and comfort limits






Existing work has suggested using clustering techniques to automatically categorize traffic scenarios or driving behaviours~\cite{mentalModels,simMatrixProfile,connectedVclustering,forestclutsering,zhao2021large}. These approaches are geared towards clustering real-time, multi-trajectory, and multivariate time series data into similar driving encounters or scenario types. Unlike these techniques, \approach aims to eliminate duplicate violations by clustering scenarios with violations. Hence, \textit{Duplicate Violations Detector} only requires a carefully-selected, smaller number of features involving just a few time frames in a scenario. 

\textit{Duplicate Violations Detector} clusters driving scenarios with similar violations into groups. For the clustering itself, we chose DBSCAN (i.e., density-based spatial clustering of applications with noise)~\cite{dbscan}, since it is more suited for spatial data. We also experimented with k-means, which resulted in clusters of undesired structure and quality. We avoided the use of hierarchical clustering~\cite{hierarchicalClustering} due to it computationally expensive nature.

\section{Evaluation}\label{sec:eval}

In order to empirically evaluate \approach, and to understand how its configuration affects the quality of generated tests, we investigate the following research questions:

\begin{enumerate}[leftmargin=*,label=\bfseries
  RQ\arabic*:,nosep,wide=0pt]
    \item How accurate are the driving scenarios generated by \approach? 
    \item How effective are \approach's generated driving scenarios at exposing AV software to safety and comfort violations? 
    \item What is the runtime efficiency of \approach's generated tests and oracles? 
    \item To what extent does \approach eliminate duplicate violations?
\end{enumerate}
\vspace{1mm}

\noindent\textbf{Experiment Settings.} Our extensive evaluation consists of executing 31,413 virtual tests on Baidu Apollo. For this reason, we conducted our experiments on three machines: 4 AMD EPYC 7551 32-Core Processor (512GB of RAM), 1 AMD EPYC 7551 32-Core Processor (256GB of RAM), and 1 AMD Opteron 64-core Processor 6376 (256GB RAM) all running Ubuntu 18.04.5. In the current implementation of \approach, we focus our efforts on testing Baidu Apollo 6.0~\cite{baidu_apollo}, an open-source and production-grade AV software system that supports a wide variety of driving scenarios and explicitly aims for both safety and rider's comfort. 
We use Apollo's simulation feature, \texttt{Sim-Control}, to simulate driving scenarios. \texttt{Sim-Control} does not simulate the \textit{control} of the ego car; instead, the ego car acts on the planning results. 

We configured \approach to generate driving scenarios for 3 high-definition maps of cities/street blocks located in California: Sunnyvale is a large map consisting of 3061 lanes, with a total length of 107 km; San Mateo is a medium map with 1,305 lanes and a total length of 24 km; and Borregas Ave, which is a small map of a city block in Sunnyvale with 60 lanes and a total length of 3 km. The three maps consist of various types of road curvature (e.g., straight, curved, intersections) and different types of lanes (e.g., highways, city roads, bike lanes, etc.). As a result, these maps are highly representative of real-world AV driving scenarios with a wide variety of diverse environmental elements. 
\looseness-1

To evaluate the effectiveness of our full approach (\approachPlus), we compare it with a state-of-the-art search-based testing approach using only a partially mutable representation (\approachMinus) and a random version of our approach that leverages \approach's \textit{domain-specific constraints} and randomly generates obstacles (\approachRand). All three representations of \approach contain the same components described in \autoref{fig:overview}, except for the \textit{Scenario Generator}, which dictates how driving scenarios are generated. \approachPlus and \approachMinus use a genetic algorithm to guide the test generation by maximizing unique violations. While \approachPlus represents obstacles as individuals allowing them to be \textit{fully mutable}, \approachMinus represents obstacles as genes, resulting in them being \textit{partially mutable}. Both \approachPlus and \approachMinus use the same search operators algorithms described in \autoref{search_op}. The  \textit{Scenario Generator} in \approachRand does not contain any genetic algorithm, and it produces driving scenarios by randomly generating obstacles. 

We evolved populations of 50 scenarios per generation, each with a minimum of 1 obstacle per scenario and a maximum of 70 obstacles. We configured the maximum scenario duration to be 30 seconds and stopped scenario generation after 12 hours. We used the crossover operator with a probability of 0.8 and mutated single individuals with a probability of 0.2. Mutating a scenario by either adding a new obstacle from another scenario or removing an obstacle was performed with a probability of 0.1 each. We run each \textit{representation} (\approachPlus, \approachMinus and \approachRand) on all maps (Borregas, San Mateo and Sunnyvale), and repeated each experiment 5 times resulting in a total of 45 experiments and 540 hours of test executions. 
\begin{figure}[ht]
\includegraphics[width=1\linewidth]{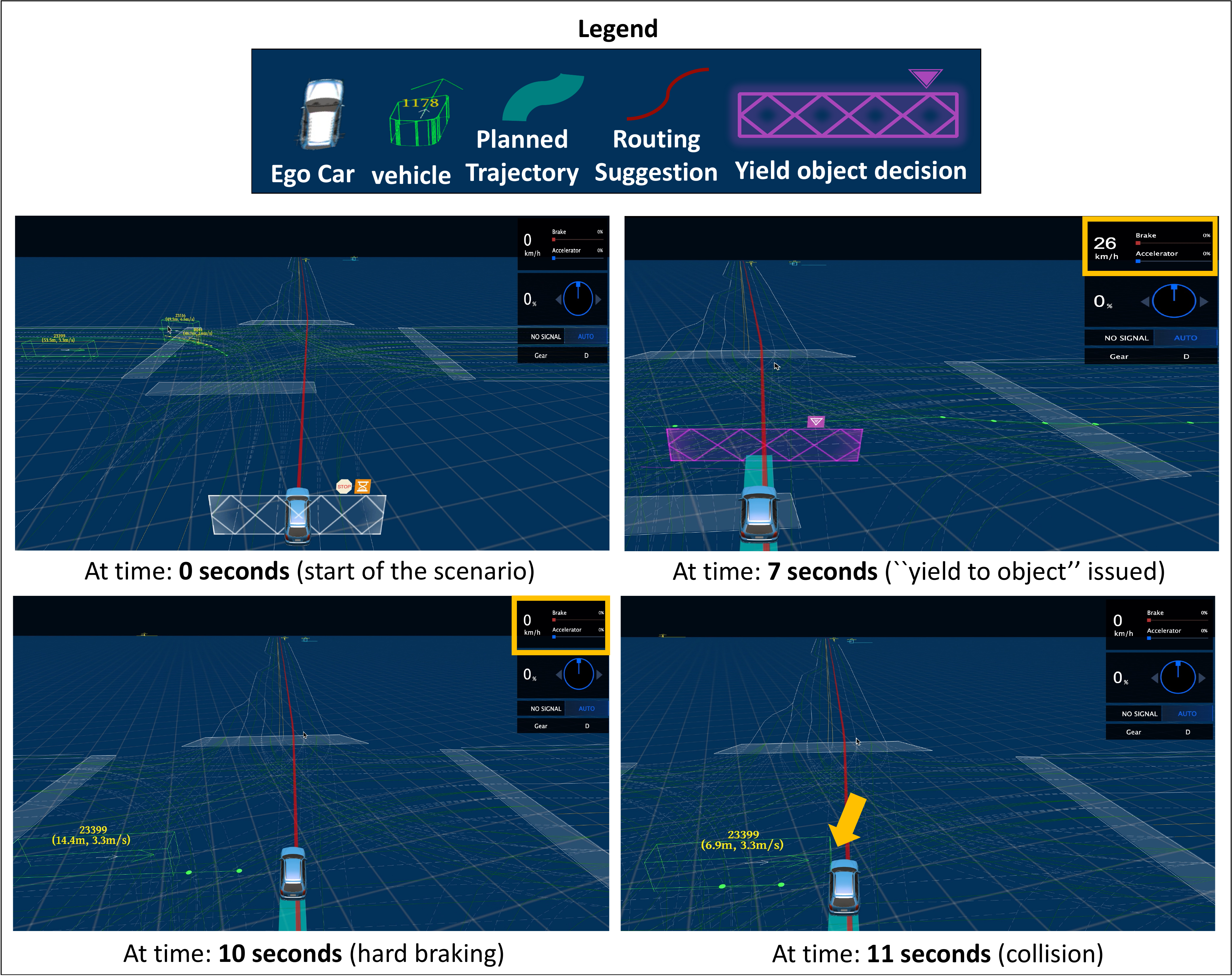}
\caption{An example of one scenario generated by \approach with two reported violations: collision and hard braking. This image is obtained from Dreamview, the visual simulator of Apollo.}
\label{fig:case-study}
\end{figure}

\subsection{RQ1: Accuracy of Generated Scenarios}\label{eval:accruacy}

We manually verify \approach's accuracy since it is not possible to verify it otherwise, due to the fact that the tests provided by Apollo’s interfaces only check whether the configuration of modules are correct. To that end, we utilize Apollo Dreamview~\cite{dreamview}, a web-based application that visualizes the ego car's driving behaviour. As shown in \autoref{fig:case-study}, Dreamview's interface allows users to view the current behaviour of the ego car along with surrounding obstacles (i.e., pedestrians, bikes, and cars), the ego car's speed in km/h and its acceleration/braking percentage (at the top-right corner of the screen), information about the lanes (i.e., speed limit imposed by lanes), etc. 

Recall from \autoref{output_recorder}, that the \textit{Planning Output Recorder} is responsible for storing the ego car's driving behaviours in record files, that can be replayed on Dreamview. Furthermore, the \textit{Grading Metrics Checker} (\autoref{sec:scenorita:gmc})---while evaluating scenarios for safety and comfort violations---collects some metadata related to such violations. For example, if a collision is reported for a scenario, we collect information related to the obstacle that collided with the ego car (i.e., $T_{O_k}$ and $ID_{O_k}$), the type of collision (i.e., rear-end, left, right), the location of the collision on the map, etc. To this end, three of the authors painstakingly and carefully evaluated the accuracy of generated scenarios by (1) replaying these scenarios on Dreamview and (2) comparing any observed violations (e.g., whether the ego car collided with obstacle $o$ in lane $l$, or if the speed of the ego car exceeded the speed limit imposed by a lane) with those reported by the \textit{Grading Metrics checker}. 

From a total of 45 experiments, where each experiment generates $\approx$400 violations, we verify $\approx$90 each (23\%). Since the manual verification takes 1.5-2 minutes per scenario, the total time to verify 23\% of all generated scenarios is 135 hours. Due to the time-consuming nature of manually verifying the accuracy of generated scenarios on Dreamview, we randomly sample a total of 4,426 out of 19,247 reported violations (23\%), and visualize them on Dreamview to verify their correctness. To keep our sample representative of all violations, we select sample scenarios with (i) a single violation (e.g., collision only), (ii) multiple violations (up to 4 per scenario), (iii) different combinations of violations per scenario (e.g., collision and speed, collision and hard braking), and (iv) scenarios selected from different times during the test generation process (i.e., scenarios were selected from the first few hours, after 6 hours, and the last few hours). All of the scenarios in our sample are accurate, i.e., the reported violations are consistent with the violations observed on Dreamview. Details of the verified scenarios are available in \cite{scenorita-repo}. As a result, we find that:

\begin{finding}
\label{find:accuracy}
Our manual verification of 23\% of reported violations show that \approach is able to generate scenarios with highly accurate safety and comfort violations. 
\end{finding}

\subsection{RQ2: Effectiveness at Producing Scenarios with Safety and Comfort Violations}\label{eval:effectiveness} 

RQ2 investigates whether using our approach (\approachPlus) leads to more reported violations, compared to random search (\approachRand) and the state-of-the-art search-based testing (\approachMinus). In conducting our evaluation of this research question, we followed the guidelines in \cite{randomTestGuide} for comparing \approachPlus against both \approachMinus and \approachRand. Hence, we performed two statistical tests: 1) \textit{Mann-Whitney U-test $p$-values} to determine statistical differences, and 2) \textit{Vargha–Delaney's $\hat{A_{12}}$ index}~\cite{varghaCritique} to determine the effect size. The results of these tests, in our experiments, are interpreted as follows: If Mann-Whitney U-test produces $p\leq0.05$, this indicates that there is a significant difference between the quality
of solutions provided by \approachPlus and \approachMinus or \approachRand. The $\hat{A_{12}}$ statistical test measures how often, on average, one approach outperforms another; if $\hat{A_{12}}=0.5$, then the two approaches achieve equal performance; if $\hat{A_{12}}>0.5$, then the first approach is better; otherwise, the first approach is worse. The closer $\hat{A_{12}}$ is to 0.5, the smaller the difference between the techniques; the further $\hat{A_{12}}$ is from 0.5, the larger the difference.

\autoref{tab:rq2_4} summarizes the number of \textit{unique} violations found, by each representation, for all maps. An exhaustive list of unique violations associated with each map is available online~\cite{scenorita-repo}. From \autoref{tab:rq2_4}, we observe that \approachPlus found, on average, a total of 1,026 unique violations in all maps over the course of our experiments. Furthermore, we find that \approachPlus discovered 23.47\% more violations compared to \approachMinus ($\hat{A_{12}}=0.84,p<0.05$), and 24.21\% more violations compared to \approachRand ($\hat{A_{12}}=0.89,p<0.05$). 


\begin{table}[!htp]\centering
\scriptsize
\begin{tabular}{L{1.1cm}C{0.43cm}C{0.43cm}C{0.43cm}C{0.43cm}C{0.43cm}C{0.43cm}C{0.43cm}C{0.43cm}C{0.43cm}}\toprule
&\multicolumn{3}{c}{\textbf{\approachPlus}} &\multicolumn{3}{c}{\textbf{\approachMinus}} &\multicolumn{3}{c}{\textbf{\approachRand}} \\\cmidrule{2-10}
\textbf{} &\textbf{All Viol.} &\textbf{Uniq. Viol.} &\textbf{Elim. (\%)} &\textbf{All Viol.} &\textbf{Uniq. Viol.} &\textbf{Elim. (\%)} &\textbf{All Viol.} &\textbf{Uniq. Viol.} &\textbf{Elim. (\%)} \\\midrule
\textbf{Collision} &411 &386 &\cellcolor{lgray}6.08\% & 328 & 246 &\cellcolor{lgray}25.00\% & 305 &264 &\cellcolor{lgray}13.44\% \\
\textbf{Speed} &25 &21 &\cellcolor{lgray}16.00\% & 24 & 18 &\cellcolor{lgray}25.00\% &27 & 18 &\cellcolor{lgray}33.33\% \\
\textbf{UnasfeLane Change} &497 &291 &\cellcolor{lgray}41.45\% &506 &275 &\cellcolor{lgray}45.65\% &509 &269 &\cellcolor{lgray}47.15\% \\
\textbf{FastAccl} &212 &132 &\cellcolor{lgray}37.74\% &183 &109 &\cellcolor{lgray}40.44\% & 188 & 109 &\cellcolor{lgray}42.02\% \\
\textbf{HardBrake} &223 &196 &\cellcolor{lgray}12.11\% & 217 &183 &\cellcolor{lgray}15.67\% & 196 &166 &\cellcolor{lgray}15.31\% \\\midrule
$\boldsymbol{\mathit{Total_{violations}}}$  &\textbf{1368} &\textbf{1,026} &\cellcolor{lgray}\textbf{25.00\%} &\textbf{1258} &\textbf{831} &\cellcolor{lgray}\textbf{33.94\%} &\textbf{1225} &\textbf{826} &\cellcolor{lgray}\textbf{32.57\%} \\
\bottomrule
\end{tabular}
\caption{The number of all violations (All Viol.) reported by \approachPlus, \approachMinus, and \approachRand, along with the total number of unique violations (Unique Viol.), and the percentage of duplicate violations eliminated (Elim. (\%)).}\label{tab:rq2_4}
\end{table}
\vspace{-1mm}

For the collision violation, \approachPlus finds, on average, 386 collisions: a 56.91\% increase compared to \approachMinus ($\hat{A_{12}}=0.80$), and a 46.21\% increase compared to \approachRand ($\hat{A_{12}}=0.77$). We observe a similar trend with the fast acceleration oracle, where \approachPlus reports 21.10\% more fast acceleration violations compared to each \approachMinus and \approachRand, respectively ($\hat{A_{12}}=0.63$ and $\hat{A_{12}}=0.61$).

\approachPlus reports a 16.67\% increase in speeding violations, on average, compared to \approachMinus and \approachRand. Furthermore, our results using the $\hat{A_{12}}$ measure, indicate that \approachPlus \textit{statistically} outperforms the latter approaches: For 57\% and 58\% of the time, \approachPlus reports more speed violations compared to \approachRand ($\hat{A_{12}}=0.57$) and \approachMinus ($\hat{A_{12}}=0.58$). Similarly, \approachPlus finds 13 more hard braking violations compared to \approachMinus, and 30 more violations compared to \approachRand;  \approachPlus reports more hard braking violations in 59\% and 71\% of the time compared to the other approaches. As for unsafe lane change, \approachPlus finds, on average, 291 violations; a 5.82\% increase compared to \approachMinus($\hat{A_{12}}=0.54$), and a 8.18\% compared to \approachRand($\hat{A_{12}}=0.54$).

The collision violations obtained with \approachPlus were significantly higher than \approachMinus and \approachRand, compared to the other violations (i.e., speeding, unsafe lane change and hard braking). This result is likely due to the fact that the collision detection encodes obstacles' behaviours in its fitness function (\autoref{fitness}). One potential way to improve the reported violations for other oracles can be achieved by encoding more elements into the fitness function of other violations. 
From these results, we find that:

\begin{finding}
\label{find:effectivness}
In our experiments, \approachPlus found a total of  \textbf{1,026} safety and comfort violations including: \textbf{386} collisions, \textbf{21} speed violations, \textbf{291} unsafe lane changes, \textbf{132} fast acceleration violations, and \textbf{196} hard braking violations. Overall \approachPlus finds, on average, \textbf{23.47\%} more violations compared to \approachMinus, and \textbf{24.21\%} more violations compared to \approachRand. 
\end{finding}

Our online dataset~\cite{scenorita-repo} contains a list of ten case studies that demonstrate \approach's ability to generate effective and valid scenarios which expose the ego car to critical situations. The case studies in~\cite{scenorita-repo} demonstrate Apollo's limited ability to handle unexpected behaviours by other obstacles (bicycles, pedestrians, and cars) such as speeding, not yielding to traffic, jaywalking, changing lanes suddenly, etc. This is alarming since (1) aggressive drivers, jaywalkers, and unexpected weather and road conditions are common in real life and (2) AV software, such as Apollo, should cope with such conditions. Space limitations preclude us from visualizing and discussing all ten cases, therefore, we discuss one case study in details (case study 9) and refer the reader to our dataset~\cite{scenorita-repo} for video recordings and details of the remaining case studies. 

\noindent\textbf{Case Study.} \autoref{fig:case-study} shows a scenario with two reported violations: \textbf{collision} and \textbf{hard braking}. The ego car is shown approaching an intersection at the beginning of the scenario (0 seconds). The ego car fails to predict the movement of another obstacle (a truck) crossing the same intersection from the left side. At 7 seconds, a decision to ``yield to an object'' (purple rectangle) is issued by the planning module, causing the ego car to brake suddenly and stop abruptly at 10 seconds; the speed of the ego car went down from 26 km/hr to 0 km/hr within less than 2 seconds, resulting in a motion sickness-inducing deceleration value of $-4.3\,m/s^2$. Since the ego car stopped in the middle of an intersection with oncoming traffic, the truck travelling on its left collides with it at 11 seconds. These two violations occur due to the planning module being unable to react to a sudden change in an obstacle behaviour. 
From these case studies, we determine the following finding:
\begin{finding}
\label{find:effectivness2}
The case studies show that \approach is capable of generating complex and effective scenarios that expose the ego car to critical and realistic situations. These case studies demonstrate Apollo's inability to cope with unexpected behaviours from obstacles, such as aggressive drivers or jaywalkers.   
\end{finding}

\subsection{RQ3: Efficiency of \approach} \label{eval:efficiency}
\begin{table}[!htp]\centering
\scriptsize
\begin{tabular}{lcccc}\toprule
&\multicolumn{4}{c}{\textbf{Execution Time (sec.)}} \\\cmidrule{2-5}
&\textbf{Simulation} &\textbf{Oracles} &\textbf{MISC} &\textbf{E2E} \\\midrule
\approachPlus &41.52 &11.97 &10.43 &63.92 \\
\approachMinus &42.09 &12.26 &10.17 &64.51 \\
\approachRand &41.60 &12.07 &9.16 &62.83 \\
\bottomrule
\end{tabular}
\caption{Efficiency of generated scenarios by \approachPlus, \approachMinus and \approachRand}\label{tab:efficiency}
\end{table}
\vspace{-5pt}
In RQ3, we study the efficiency of \approachPlus by measuring its execution time, and comparing it with the execution time of \approachMinus and \approachRand. \autoref{tab:efficiency} shows that \approachPlus takes \textbf{63.92} seconds, on average, to execute a scenario from end-to-end (\textit{E2E}); it takes \textbf{10.43} seconds, on average, to generate the scenario representation and confirm its validity according to the \textit{domain-specific constraints} (\textit{MISC}); \textbf{41.52} seconds to generate the corresponding driving simulation (\textit{Simulation}); and \textbf{11.97} seconds for checking the grading metrics (\textit{Oracles}). Simulating driving scenarios is time-consuming (e.g., transforming the scenario representation into simulations, running each simulation for 30 seconds, and recording the car behaviour); hence, the scenario simulation stage strongly affects the efficiency of the overall test generation process in \approach. We further observe that the difference in execution-time between all three representations (\approachPlus, \approachMinus and \approachRand) is negligible. From these results, we find that:
\begin{finding}
\label{find:effeciency}
\approachPlus is efficient, with an average runtime of 63.92 seconds per scenario, and can be used in practice to generate driving scenarios that expose AV software to safety violations. Moreover, \approachPlus managed to generate 23.47\% and 24.21\% more violations compared to the two other representation in the same amount of time.  
\end{finding}

\subsection{RQ4: Duplicate Violation Detection}\label{eval:dvd}

This RQ investigates the extent to which \approach eliminates similar violations, and compares the percentage of duplicate violations generated by all three representations (\approachPlus, \approachMinus and \approachRand). To answer this RQ, we configure DBSCAN~\cite{dbscan} to cluster the scenarios with similar violations into the same group, based on a set of features as described in \autoref{sec:scenorita:dvd}. We adopted the approach in \cite{dbscan-eps} to automatically determine the optimal value for \textit{epsilon}; \textit{epsilon} defines the maximum distance allowed between two points within the same cluster. Eliminating duplicate violations is quick and takes, on average, 0.1 seconds per experiment.  
  
To confirm the correctness of generated clusters, three of the authors manually and independently evaluated the accuracy of generated clusters in 18 randomly-selected experiments out of a total of 45 (40\%). The authors examined violations in the same clusters to confirm whether they are similar by comparing a set of features associated with each scenario in the cluster. For example, consider two scenarios, \texttt{Scenario1} and \texttt{Scenario17}, both of which are in the same cluster and have a collision violation: In such a case, we compare the set of features (from \autoref{sec:scenorita:dvd}) of the two scenarios to confirm that the collision occurred in the same position in \texttt{Scenario1} as it did in \texttt{Scenario17} ($p^{E}_{t_c}$), the ego car in both scenarios collided with an obstacle with the same type ($T_{O_k}$) and size ($Z_{O_k}$), the crash in both scenarios is the same ($collision^{type}$), etc. The authors also replayed these scenarios on Dreamview to observe if the scenarios in one cluster have similar violations.

\autoref{tab:rq2_4} shows all violations (including duplicates) generated by \approachPlus, \approachMinus, and \approachRand along with the unique number of violations (generated by the \textit{Duplicate Violations Detector}), and the percentage of eliminated violations. An exhaustive list of eliminated violations based on each map is available in our online dataset~\cite{scenorita-repo}. From the results in \autoref{tab:rq2_4}, we observe that \approachPlus eliminated, on average, a total of 342 similar violations in all maps over the course of our experiments. 
Overall, we find that:
\begin{finding}
\label{find:effeciency}
Our manual verification of 40\% of our experiments shows that \approach is able to identify and eliminate duplicate tests. The \textit{Duplicate Violations Detector} eliminated 25\% duplicate tests in \approachPlus, 33.94\% in \approachMinus, and 32.57\% in \approachRand---indicating that \approachPlus generates more unique violations compared to the two other representations.
\end{finding}



\section{Threats to Validity}\label{sec:threats}
\textbf{Internal threats.} One potential threat to internal validity is the selection of scenario durations: Simulation-based tests require the execution of time-consuming computer simulations to produce violations. 
We determined from our experimentation that our selected scenario duration of 30 seconds finds a significant number and variety of violations without incurring drastically long test execution times.

To account for validity threats arising from randomness in search algorithms, we follow the guidelines in \cite{randomTestGuide}: (i) we repeated the experiments for each representation (\approachPlus, \approachMinus, and \approachRand) 15 times, (ii) we used the non-parametric Mann-Whitney U-test to detect statistical differences and reported the obtained $p$-value, and (iii) we reported $\hat{A_{12}}$ index (a standardized effect size measure). We make our full experimental results available in \cite{scenorita-repo}.

To mitigate threats arising from our selection of search operators, we selected (i) a widely-used algorithm in the search-based software engineering (SBSE) community, i.e., NSGA-II and (ii) crossover and mutation algorithms that best fits our gene representation. For parameter tuning, we followed the guidelines in~\cite{SBSEguidlines,randomTestGuide}---which suggests that 
standard parameter settings are usually
recommended---leading us to use default settings in DEAP-1.3 ~\cite{deap}, the framework used in our search-based implementation. 

\textbf{External threats.} One external threat is that we applied \approach to a single AV software system, Apollo. To mitigate the threat, we selected the only high autonomy (i.e., Level 4), open-source, production-grade AV software system that supports a wide variety of driving scenarios and explicitly aims for both safety and driver comfort. To mitigate threats related to generalizability of our results to other maps, we applied \approach to three high-definition maps of cities in California: Borregas (60 lanes), San Mateo (1,305 lanes), and Sunnyvale (3,061 lanes). Note that Autoware~\cite{autoware}, despite being open-source and widely-used~\cite{autoware_widely_used}, is considered a research-grade and not a production-grade AV software system~\cite{kato2015autoware,kato2018autoware}, which we further verified through speaking with Christian John, the Vice Chair and Chief Software Architect of Autoware. 

\textbf{Construct Validity} The main threat to construct validity is how we measure and calculate safety and comfort violations. To mitigate this threat, we measure these violations using grading metrics defined by Apollo's developers~\cite{grading_metrics}. We utilize thresholds (e.g., speeding or acceleration thresholds) set by Apollo's developers \cite{grading_metrics}; the U. S. Department of Transportation \cite{lanechange_duration}; or thresholds used by major AV companies (e.g., Alphabet Waymo \cite{speed_limit_google}). 

\section{Related Work}

A wide array of studies focus on applying traditional testing techniques to AVs including adaptive stress testing~\cite{corso2019adaptive}, where noise is injected into the input sensors of an AV to cause accidents; fitness function templates for testing automated and autonomous driving systems with heuristic search \cite{hauer2019fitness}; and search-based optimization \cite{kluck2019genetic}. These studies provide limited insights into the testing of real-world AVs, since they do not evaluate their techniques on open-source,  production-grade AV software.

Other related work reproduce tests from real crashes such as \cite{erbsmehl2009simulation}, where car crashes are recreated by replaying the sensory data collected during physical-world crashes. Similarly, \textit{AC3R}~\cite{ac3r} generates driving simulations which reproduces car crashes from police reports using natural language processing (NLP). However, \textit{AC3R} requires manual collection of police reports and inherit the accuracy limitations of the underlying NLP used to extract information from police reports. 

Another line of work, combines procedural content generation and search-based testing to automatically create challenging virtual scenarios for AV software \cite{asfault}. However, \textit{AsFault}, does not take into account the behaviour of other obstacles when testing for safety violations in AVs. Calò et al. \cite{avoidCollision} proposed two search-based approaches for finding \textit{avoidable} collisions. They define comfort and speed as weights to rank short-term paths. However, they do not formalize comfort and speed in the fitness function, nor do they evaluate them. Furthermore, their work is applied to closed-source anonymous software, with no information about the maps used or scenarios’ configurations. 

\textit{AV-Fuzzer}~\cite{avfuzzer} perturbs driving maneuvers of traffic participants to create situations in which an AV can run into safety violations. However, \textit{AV-Fuzzer} requires the user to \emph{manually} define a \textit{driving environment} by specifying limited maneuvers (i.e., only acceleration/deceleration, lane following, and lane changing) of obstacles. \approach \emph{automatically} generates the maneuvers of obstacles and even supports a wider variety of maneuvers: lane follow, lane change, left turn, right turn, U-turn, and lane merge.

A series of work focuses on the vision and machine-learning aspects of AV software~\cite{deepbillboard,stocco2020misbehaviour,deeproad,kim2020reducing,peng2020first,abdessalem2018testing,ben2016testing}. Rather than focus on these aspects, \approach targets the planning component of AV software. Previous work has shown that the most bug-ridden component of production-grade, open-source AV software systems is the planning component as opposed to the components responsible for or utilizing vision or machine-learning capabilities of AVs~\cite{AVbugStudy}. 

Unlike previous work, the gene representation in \approach allows the states and properties of obstacles in a scenario to be \textit{fully mutable}. Previous techniques either do not specify their gene representations or only allow obstacles to be \textit{partially mutable}, avoiding having to deal with generating valid obstacles trajectories. The aforementioned previous work uses highly limited numbers of oracle types and fitness functions---mainly consisting of one test oracle or fitness function per work (either collision detection or lane keeping) and ignoring motion sickness-inducing violations. \approach, however, utilizes 3 safety oracles (i.e., collision detection, speeding detection, and unsafe lane change), 2 motion sickness-oriented oracles (i.e., fast acceleration and hard braking detection), and includes associated fitness functions. The lack of sufficient test oracles and fitness functions in previous work reduces the safety assurance and ignores vehicle-induced motion sickness---which many people (e.g., one-third of Americans) suffer from in the world~\cite{nyt_motion_sickness,michigan_motion_sickness,leung2019motion}.

\section{Conclusion}
In this paper, we propose \approach, a novel search-based testing framework, which generates AV software to 3 types of safety-critical and 2 types of motion sickness-inducing scenarios in a manner that reduces duplicate scenarios, allows fully mutable obstacles with valid and modifiable obstacles trajectories, and follows domain-specific constraints obtained from authoritative sources.  We evaluate \approach on Baidu Apollo, a high autonomy (L4), open-source, and production-grade AV software system that supports a wide variety of driving scenarios. We compare our approach (\approachPlus) with a state-of-the-art search-based testing approach using only a partially mutable representation (\approachMinus) and a random version of our approach that leverages \approach's \textit{domain-specific constraints} and randomly-generated obstacles.  \approachPlus found a total of 1,026 unique safety and comfort violations including: 386 collisions, 21 speed violations, 291 unsafe lane changes, 132 fast acceleration violations, and 196 hard-braking violations. Moreover, \approachPlus generates, on average, 23.47\% and 24.21\% more violations compared to the two other representations in the same amount of time (63.92 sec/scenario). For future work, we aim to expand \approach to handle (i) generation of scenarios and oracles focused on traffic lights and stop signs and (ii) extending the work to other AV software systems (e.g., Autoware).
\clearpage
\bibliographystyle{ACM-Reference-Format}
\bibliography{main}

\end{document}